\documentclass[prb,aps,twocolumn]{revtex4}

\usepackage{amsmath}
\usepackage{diagbox}

\usepackage[dvips]{graphicx}

\newcommand{\ket}[2][]{{|#2\rangle_{#1}}}
\newcommand{\bra}[2][]{{}_{#1}\langle #2|}
\newcommand{\braket}[3][]{{{}_{#1}\langle#2|#3\rangle_{#1}}}
\newcommand{\proj}[2][]{\ket{#2}_{#1}\bra{#2}}
\newcommand{\Tr}{\text{Tr}}

\begin{document}

\title{Quantum mechanical which-way experiment with an internal degree of freedom}
\author{Konrad Banaszek,$^1$ Pawe{\l} Horodecki,$^{2,3}$ Micha{\l} Karpi\'{n}ski,$^1$}
\thanks{Current address: Clarendon Laboratory, University of Oxford, Parks Road, Oxford, OX1 3PU, UK}
\author{Czes{\l}aw Radzewicz$^1$}
\affiliation{$^1$Faculty of Physics, University of Warsaw, ul.\ Ho\.{z}a 69, PL-00-681 Warszawa, Poland}
\affiliation{$^2$Faculty of Applied Physics and Mathematics, Technical University of Gda\'{n}sk, ul.\ Narutowicza 11/12, PL-80-952 Gda\'{n}sk, Poland}
\affiliation{$^3$National Quantum Information Center of Gda\'{n}sk, ul. W{\l}. Andersa 27, PL-81-824 Sopot, Poland}

\begin{abstract}
For a particle travelling through an interferometer, the trade-off between the available which-way information and the interference visibility provides a lucid manifestation of the quantum mechanical wave-particle duality.
Here we analyze this relation for a particle possessing an internal degree of freedom such as spin. We quantify the trade-off with a general inequality that paints an unexpectedly intricate picture of wave-particle duality when internal states are involved. Strikingly, in some instances which-way information becomes erased by introducing classical uncertainty in the internal degree of freedom.
Furthermore, even imperfect interference visibility measured for a suitable set of spin preparations can be sufficient to infer absence of which-way information. General results are illustrated with a proof-of-principle single photon experiment.
\end{abstract}

\date{\today}
\maketitle

The duality between wave and particle properties of a microscopic physical system is a founding principle of
quantum mechanics.\cite{Bohr1,Bohr2} A canonical illustration is provided by a single particle travelling through a double slit or a Mach-Zehnder interferometer:
an attempt to gain information about the path taken by the particle inevitably reduces the visibility of the interference pattern.\cite{WoottersZurek,ScullyEnglertWalther,GreenbergerYasin,Englert,JaegerShimony,BjorkKarlsson,JacquesScience,PeruzzoScience,KaiserScience}
The purpose of this work is to analyze the trade-off between interference visibility and which-way information for a particle equipped with an internal degree of freedom, for example spin. In such a general case, the interaction with an environment that acquires which-way information can transform in a non-trivial manner the joint path-spin state of the particle. This opens up an interesting question of how to infer the amount of which-way information deposited in the environment from visibility measurements. Another issue is, to what extent manipulating the spin subsystem can control the information about the path taken by the particle.

A common approach to quantify the amount of which-way information is to use distinguishability $D$, defined as the maximum difference between the probabilities of correct and incorrect identification\cite{Helstrom} of the path taken by the particle inside the interferometer, based on the state of the environment. We assume here that both paths are equiprobable. In the spinless case, the ability for the particle distributed between two paths to interfere is characterized by the visibility $V$, which measures the modulation depth of interference fringes after the paths are combined at the interferometer exit. These two quantities are related by the inequality\cite{Englert}
\begin{equation}
\label{Eq:Spinless}
D^2 + V^2 \le 1.
\end{equation}

When the particle travelling through the interferometer has an internal structure, the amount of available which-way information depends in principle on the preparation of the spin subsystem and the specifics of the interaction with the environment. Here we demonstrate that in this case the strongest bound on the distinguishability $D$ is obtained by replacing $V$ with a quantity named generalized visibility that depends on the initial spin preparation and the effective quantum channel experienced by the particle resulting from the interaction with the environment. Further, we present a systematic method to construct estimates for the generalized visibility based on directly measurable quantities, i.e.\ interference visibilities for particular spin preparations at the input and selections of spin states before combining the paths at the interferometer exit. This provides an efficient strategy to find an upper bound on the available which-way information without performing full quantum process tomography.\cite{Tomo1,Tomo2,Tomo3} We illustrate the general results with a proof-of-principle single photon experiment that also demonstrates how which-way information can be erased by introducing mixedness in the spin preparation.

\section*{Results}

{\bf The inequality.}
The quantum mechanical system considered here is a particle present in one of two distinguishable paths,
denoted respectively as $\ket[Q]{0}$ and $\ket[Q]{1}$. Additionally, the particle is equipped with a $d$-dimensional internal degree of freedom---spin---which we will treat as a subsystem $S$. The complete state of the particle prepared in the interferometer is characterized by a certain density operator $\hat{\varrho}_{QS}$.
Both paths are equiprobable, i.e.\ $\Tr_{S}\bigl( \bra[Q]{0} \hat{\varrho}_{QS} \ket[Q]{0} \bigr) = \Tr_{S}\bigl( \bra[Q]{1} \hat{\varrho}_{QS} \ket[Q]{1} \bigr) = \frac{1}{2}$.
Inside the interferometer, the particle interacts with an environment $E$ producing a joint state $\hat{\varrho}'_{QSE}$. The interaction is constrained in that it cannot transfer the particle between the paths. In this setting,
which-way information contained in the environment corresponds to the ability to discriminate between two normalized density matrices $\mbox{$\hat{\varrho}'$}_{E}^{(i)} = 2 \Tr_{S} \bigl( \bra[Q]{i} \hat{\varrho}'_{QSE} \ket[Q]{i}\bigr)$, $i=0,1$.
In the general case of mixed states, distinguishability is given explicitly by the expression\cite{Helstrom,Fuchs}
\begin{equation}
\label{Eq:Distinguishability}
D = \frac{1}{2} || \mbox{$\hat{\varrho}'$}_{E}^{(0)} - \mbox{$\hat{\varrho}'$}_{E}^{(1)}||,
\end{equation}
where $|| \cdot ||$ denotes the trace norm. Let us stress here that this quantity is not conditioned upon any measurement performed eventually on subsystems $QS$.

Our aim is to infer an upper bound on $D$ from the properties of the particle, i.e.\ subsystems $QS$, after the interaction with the environment.
The state of the particle at this stage can be written as a result of an action of a certain quantum channel\cite{NielsenChuang} $\boldsymbol\Lambda$ on the input state $\hat{\varrho}_{QS}$, with ${\boldsymbol\Lambda}(\hat{\varrho}_{QS}) = \Tr_E(\hat{\varrho}'_{QSE})$. Because the particle is not transferred between interferometer arms, for any spin operator $\hat{\sigma}_{S}$ we have
\begin{equation}
\label{Eq:Lambda01}
\boldsymbol\Lambda \big( \ket[Q]{i}\bra{j} \otimes \hat{\sigma}_{S} \bigr) = \ket[Q]{i}\bra{j} \otimes \boldsymbol\Lambda_{ij}(\hat{\sigma}_{S}),
\quad i,j=0,1.
\end{equation}
Our central result is that $D$ satisfies the inequality:
\begin{equation}
\label{Eq:DandV}
D^2 + V_G^2 \le 1,
\end{equation}
where $V_G$, named generalized visibility, reads:
\begin{equation}
\label{Eq:GenVis}
V_G = d \bigl|\bigl| ( {\bf I} \otimes \boldsymbol\Lambda_{01}) \bigl( ( \hat{\openone} \otimes \sqrt{\hat{\varrho}_0}) \proj{\Phi_+}
( \hat{\openone} \otimes \sqrt{\hat{\varrho}_1}) \bigr) \bigr|\bigl|.
\end{equation}
Here $\ket{\Phi_+}$ is a normalized maximally entangled state of two replicas of the spin subsystem, ${\bf I}$ is the identity channel, $\hat{\openone}$ is the identity operator on the spin subsystem, and $\hat{\varrho}_{i} = 2 \bra[Q]{i} \hat{\varrho}_{QS} \ket[Q]{i}$ are normalized spin states for the particle in either arm of the interferometer before the interaction with the environment. The complete proof of the inequality is presented in Methods.

{\bf Examples.}
The inequality (\ref{Eq:DandV}) resembles that for a spinless particle given in Eq.~(\ref{Eq:Spinless}), but the concept of generalized visibility covers a much wider range of physical scenarios. In order to illustrate its breadth, let us consider here three cases. First, when no interaction with the environment occurs, $\boldsymbol\Lambda_{01} = {\bf I}$. In this case $V_G = \sqrt{\Tr\hat{\varrho}_0\Tr\hat{\varrho}_1} = 1$, i.e.\ no matter what internal states for individual paths are initially chosen, the environment does not gain any which-way information.

Second, let us take $\boldsymbol\Lambda_{01}(\hat{\sigma}) = \hat{\sigma}_0 \Tr\hat{\sigma}$, where $\hat{\sigma}_0$ is a fixed unit-trace hermitian operator. Physically, this occurs when the internal degree of freedom is entirely transferred to the environment and replaced with a state $\hat{\sigma}_0$. Then $V_G = ||\sqrt{\hat{\varrho}_0} \sqrt{\hat{\varrho}_1} ||$ which is the standard expression for quantum fidelity\cite{Uhlmann,Jozsa} between density operators $\hat{\varrho}_0$ and $\hat{\varrho}_1$. Consequently, the only way to ensure that no which-way information leaks out is to associate both the interferometer paths
with identical internal states.

Finally, suppose that
\begin{equation}
\label{Eq:Lambda01transp}
{\boldsymbol\Lambda_{01}}(\hat{\sigma}) = \frac{1}{d} \hat{\sigma}^T,
\end{equation}
where $^T$ denotes transposition. This channel gives $V_G = \frac{1}{d}||\sqrt{\hat{\varrho}_0}|| \, ||\sqrt{\hat{\varrho}_1}||$. The resulting expression has non-intuitive properties: for a particle prepared in any pure state $\bigl(\ket[Q]{0} \ket[S]{\psi_0} +  \ket[Q]{1} \ket[S]{\psi_1}\bigr)/\sqrt{2}$ the generalized visibility equals $1/d$, which is strictly less than one for a non-trivial spin subsystem when $d \ge 2$. Generalized visibility reaches one only when the two paths are associated with maximally mixed spin states, $\hat{\varrho}_0 = \hat{\varrho}_1 = \frac{1}{d} \hat{\openone}$, which follows immediately from the inequality between arithmetic and quadratic means. This indicates that which-way information deposited in the environment may be erased by introducing classical noise in the input state of the spin subsystem. To illuminate this point, let us take a single photon traveling through a Mach-Zehnder interferometer and consider its polarization, spanned by two orthogonal states $\ket[S]{h}$ and $\ket[S]{v}$, as the two-dimensional internal degree of freedom. Let the interaction of the photon with the environment be given by a unitary map
\begin{widetext}
\begin{multline}
\frac{1}{\sqrt{2}}\bigl( \ket[Q]{0} \ket[S]{\psi_0} + \ket[Q]{1} \ket[S]{\psi_1} \bigr) \ket[E]{e_1} \rightarrow
\frac{1}{2} \bigl[ \ket[Q]{0} \bigl( \psi_{0h} \ket[S]{h} \ket[E]{e_1} + \psi_{0h} \ket[S]{v} \ket[E]{e_2}
+ \psi_{0v} \ket[S]{h} \ket[E]{e_3} + \psi_{0v} \ket[S]{v} \ket[E]{e_4} \bigr) \\
+ \ket[Q]{1} \bigl( \psi_{1h} \ket[S]{h} \ket[E]{e_1} + \psi_{1v} \ket[S]{h} \ket[E]{e_2}
+ \psi_{1h} \ket[S]{v} \ket[E]{e_3} + \psi_{1v} \ket[S]{v} \ket[E]{e_4} \bigr) \bigr]
\label{Eq:ChannelExplicit}
\end{multline}
\end{widetext}
where $\ket[E]{e_n}$, $n=1,\ldots, 4$, are four orthogonal states of the environment and $\psi_{ih} = \braket[S]{h}{\psi_i}$, $\psi_{iv} = \braket[S]{v}{\psi_i}$ are the probability amplitudes in the rectilinear basis. It is easy to verify that this map indeed induces ${\boldsymbol\Lambda_{01}}(\hat{\sigma}) = \hat{\sigma}^T/2$, while the polarization state in an individual arm is scrambled to a completely mixed state, ${\boldsymbol\Lambda_{00}}(\hat{\sigma}) =
{\boldsymbol\Lambda_{11}}(\hat{\sigma}) = \Tr (\hat{\sigma}) \hat{\openone}/2$.

If we prepare $\ket[S]{\psi_0} = \ket[S]{\psi_1} = \ket[S]{h}$, the two states of the environment correlated with the paths: \begin{align}
\mbox{$\hat{\varrho}'$}_{E}^{(0)} & = \frac{1}{2} \bigl( \proj[E]{e_1} + \proj[E]{e_2}\bigr) \nonumber \\  \mbox{$\hat{\varrho}'$}_{E}^{(1)} & = \frac{1}{2} \bigl( \proj[E]{e_1} + \proj[E]{e_3}\bigr)
\end{align}
are partly distinguishable.
Similarly, for the vertical input polarization, $\ket[S]{\psi_0} = \ket[S]{\psi_1} = \ket[S]{v}$, the environment states are
\begin{align}
\mbox{$\hat{\varrho}'$}_{E}^{(0)} & =  \frac{1}{2} \bigl( \proj[E]{e_3} + \proj[E]{e_4}\bigr) \nonumber \\ \mbox{$\hat{\varrho}'$}_{E}^{(1)} & =  \frac{1}{2} \bigl( \proj[E]{e_2} + \proj[E]{e_4}\bigr).
\end{align}
However, if the input polarization is prepared in a completely mixed state
$\frac{1}{2} \bigl( \proj[S]{h} + \proj[S]{v} \bigr)$,
one sees immediately that both the paths are associated with the same mixed environment state $\frac{1}{4} \sum_{k=1}^{4} \proj[E]{e_k}$ and which-way information is not available anymore. The inequality (\ref{Eq:DandV}) demonstrates that this effect is an intrinsic feature of the channel under consideration rather than its specific realization given in Eq.~(\ref{Eq:ChannelExplicit}).

{\bf Estimating generalized visibility.}
Generalized visibility $V_G$ is given by a rather intricate expression involving 
the spin preparation and the effective quantum channel experienced by the particle. In principle, full information about the channel can be obtained from quantum process tomography,\cite{Tomo1,Tomo2,Tomo3} but this approach may be resource consuming, especially for a high dimension of the internal subsystem $S$. We will now give a recipe to construct estimates for $V_G$ from direct visibility measurements for specific spin preparations and selections of individual spin components before interfering the particle paths.

Consider the following procedure. The particle is prepared in one of states
\begin{equation}
\ket[QS]{\psi^\mu} = \frac{1}{\sqrt{2}}\bigl( \ket[Q]{0} \ket[S]{\psi_0^{\mu}} +  \ket[Q]{1} \ket[S]{\psi_1^{\mu}}\bigr),
\end{equation}
where the normalized kets $\ket[S]{\psi_0^{\mu}}$ and $\ket[S]{\psi_1^{\mu}}$ describe the spin state in the upper and lower interferometer path. The index $\mu$ labels different preparations. After the interaction with the environment, the spin subsystem is filtered  individually in each interferometer arm to extract probability amplitudes corresponding to normalized spin states $\ket[S]{\chi_0^{\nu}}$ and $\ket[S]{\chi_1^{\nu}}$. These components are subsequently made indistinguishable by a suitable spin transformation and interfered on a balanced beam splitter with a relative phase shift $\phi$. The probabilities of detecting the particle at the two output ports $\pm$ of the beam splitter read
\begin{equation}
\label{Eq:p_pm}
p^{\mu\nu}_{\pm}(\phi) = \frac{1}{2} [p^{\mu\nu} \pm \text{Re}(V^{\mu\nu} e^{i\phi})].
\end{equation}
Here
\begin{equation}
p^{\mu\nu} = \frac{1}{2} \bigl[ \bra{\chi^\nu_0} {\boldsymbol\Lambda}_{00} \bigl( \proj{\psi^\mu_0} \bigr) \ket{\chi^\nu_0} +
\bra{\chi^\nu_1} {\boldsymbol\Lambda}_{11} \bigl( \proj{\psi^\mu_1} \bigr) \ket{\chi^\nu_1} \bigr]
\end{equation}
is the overall probability of detecting the particle in the filtered components. The modulation depth of interference fringes is characterized by the fractional visibility, given explicitly by
\begin{equation}
\label{Eq:FracVis}
V^{\mu\nu} = \bra{\chi_0^{\nu}} {\boldsymbol \Lambda}_{01} \bigl( \ket{\psi_0^{\mu}} \bra{\psi_1^{\mu}} \bigr) \ket{\chi_1^{\nu}}.
\end{equation}
For the quantities introduced here, we always have $|V^{\mu\nu}| \le p^{\mu\nu}$. Full-depth modulation of fringes is observed for the equality sign.

Let us now take any set of complex coefficients $\alpha_{\mu\nu}$ such that
a linear combination of rank-one
operators $\bigl( \ket{\psi_0^{\mu}}\bra{\psi_1^{\mu}} \bigr)^T \otimes \ket{\chi_1^{\nu}} \bra{\chi_0^{\nu}}$ can be written as
\begin{equation}
\label{Eq:sumklalphas}
\sum_{\mu\nu} \alpha_{\mu\nu} \bigl( \ket{\psi_0^{\mu}} \bra{\psi_1^{\mu}} \bigr)^T \otimes \ket{\chi_1^{\nu}} \bra{\chi_0^{\nu}}
= (\sqrt{\hat{\varrho}_1^T} \otimes \hat{\openone} )
\hat{U}
(\sqrt{\hat{\varrho}_0^T} \otimes \hat{\openone} )
\end{equation}
for a certain operator $\hat{U}$ acting on the duplicated spin subsystem that satisfies $\hat{U}^\dagger \hat{U} \le \hat{\openone} \otimes \hat{\openone}$. We show in Methods that under these assumptions the generalized visibility is bounded by a linear combination of fractional visibilities with the same coeffcients $\alpha_{\mu\nu}$:
\begin{equation}
\label{Eq:GenVisBound}
V_G \ge \left| \sum_{\mu\nu} \alpha_{\mu\nu} V^{\mu\nu} \right|.
\end{equation}
Combining this result with Eq.~(\ref{Eq:DandV}) yields a family of bounds on distinguishability in the form
\begin{equation}
D \le \sqrt{1-V_G^2} \le \sqrt{1- \left | \sum_{\mu\nu} \alpha_{\mu\nu} V^{\mu\nu}\right|^2} .
\label{Eq:Dlesum}
\end{equation}

\begin{figure}
\includegraphics[width=0.48\textwidth]{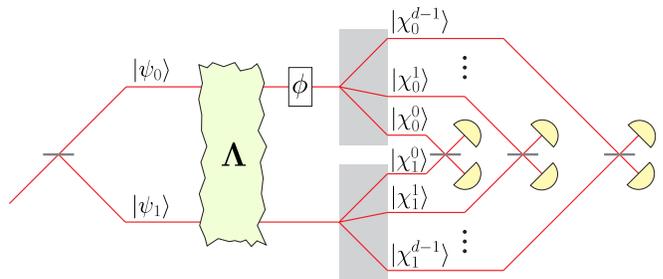}
\caption{{\bf Projective measurement of fractional visibilities.} A particle is prepared in the spin state
$\ket{\psi_0}$ in the upper arm and $\ket{\psi_1}$ in the lower arm. After interaction with the environment described by the channel
$\boldsymbol\Lambda$ the phase between interferometer arms is modulated by $\phi$. Next, the spin states are filtered in the orthonormal bases $\{\ket{\chi_0^\nu}\}$ for the upper arm and $\{\ket{\chi_1^\nu}\}$ for the lower arm, and individual components are interfered pairwise to determine fractional visibilities $V^\nu, \nu=0,1,\ldots,d-1$.}
\label{Fig:SinglePreparation}
\end{figure}

Special cases of Eq.~(\ref{Eq:Dlesum}) are intuitive.
Consider a scenario shown in Fig.~\ref{Fig:SinglePreparation}. We use only a single preparation $\bigl( \ket[Q]{0}\ket[S]{\psi_0} + \ket[Q]{1}\ket[S]{\psi_1} \bigr)/\sqrt{2}$ (henceforth in this example we drop the superfluous label $\mu$), while the sets of filter states in the upper $\{\ket[S]{\chi^\nu_0}\}$ and lower $\{\ket[S]{\chi^\nu_1}\}$ arm form two orthonormal bases for the spin subsystem with $\nu$ indexing the basis elements. In Eq.~(\ref{Eq:sumklalphas}) we can take arbitrary phase factors as coefficients $\alpha_\nu = e^{i\theta_\nu}$, with the operator $\hat{U} = \bigl( \ket{\psi_0}\bra{\psi_1} \bigr)^T \otimes \sum_{\nu} e^{i\theta_\nu} \ket{\chi_1^\nu} \bra{\chi_0^\nu}$ satisfying the condition $\hat{U}^\dagger \hat{U} \le \hat{\openone} \otimes \hat{\openone}$. According to Eq.~(\ref{Eq:GenVisBound}) we have $V_G \ge \bigl| \sum_{\nu} e^{i\theta_\nu} V^\nu \bigr|$, which maximized over phases $\theta_\nu$
estimates the generalized visibility by a sum of absolute values of fractional visibilities:
\begin{equation}
\label{Eq:VisOrthogonal}
V_G \ge \sum_{\nu} |V^\nu|.
\end{equation}
If for each filter we measure the maximum possible fractional visibility $|V^\nu|=p^\nu$, the right hand side of (\ref{Eq:VisOrthogonal}) becomes one, because completeness of the bases $\{\ket[S]{\chi^\nu_0}\}$ and $\{\ket[S]{\chi^\nu_1}\}$ implies that $\sum_{\nu} p^\nu = 1$. In other words, when the interaction with the environment introduces different phase shifts between distinguishable components of the spin subsystem but maintains overall coherence between the paths, we can observe full visibility by suitably sorting spin components at the output and interfering them pairwise. This suffices to verify that no which-way information is deposited in the environment. However, as noted earlier the interplay between the path and the spin subsystems can exhibit more intricate behaviour. We illustrate this point with a single photon experiment.

\begin{figure}
\begin{flushleft}
{\bf a)}
\end{flushleft}
\includegraphics[width=0.48\textwidth]{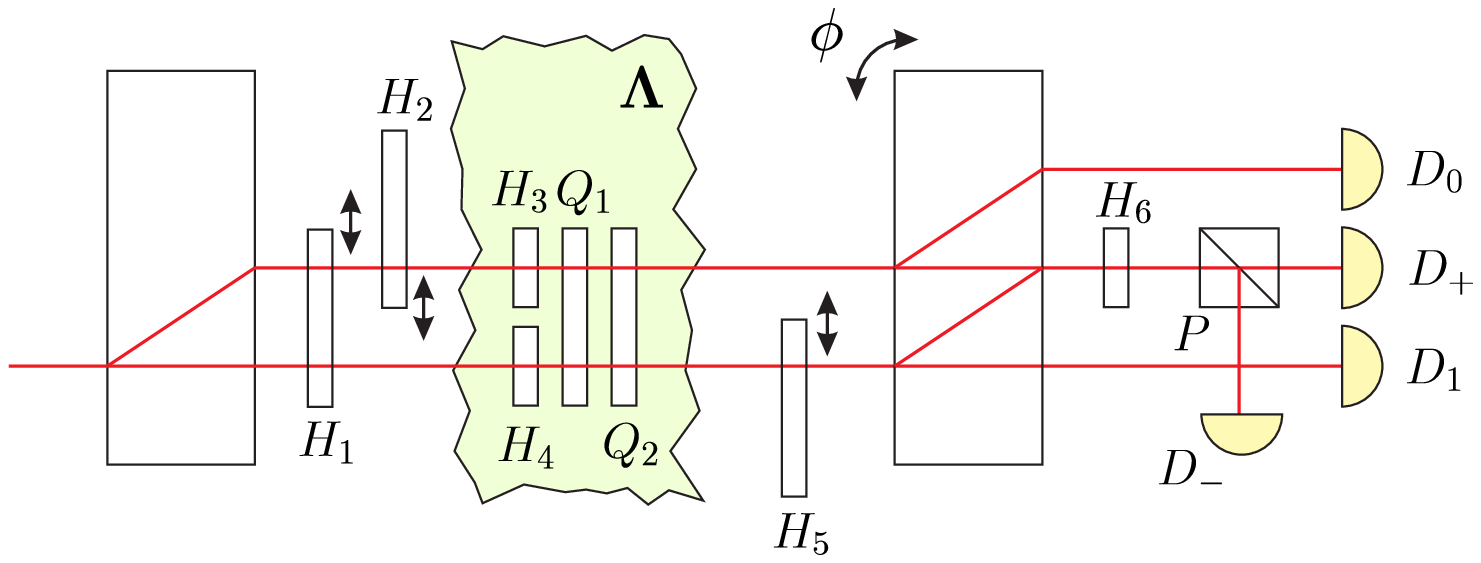}

\vspace*{5mm}

\begin{flushleft}
{\bf b)}
\end{flushleft}
\begin{tabular}{c|rrrr}
$(\hat{K}^{(0)}_S, \hat{K}^{(1)}_S)$ &
\multicolumn{1}{c}{$H_3$} & \multicolumn{1}{c}{$H_4$} & \multicolumn{1}{c}{$Q_1$} & \multicolumn{1}{c}{$Q_2$} \\
\hline
$(\hat{\openone}, \hat{\openone})$ & $-45^\circ$ & $-45^\circ$ & \phantom{$-$}$45^\circ$ & $45^\circ$ \\
$(\hat{X}, \hat{X})$ & $45^\circ$ & $45^\circ$ & $45^\circ$ & $-45^\circ$ \\
$(\hat{Y}, -\hat{Y})$ & $0^\circ$ & $90^\circ$ & $45^\circ$ & $45^\circ$ \\
$(\hat{Z}, \hat{Z})$ & $0^\circ$ & $0^\circ$ & $45^\circ$ & $-45^\circ$ \\
\end{tabular}
\caption{{\bf Noisy Mach-Zehnder interferometer.} {\bf a}, Experimental setup. A heralded single photon with 830 nm central wavelength and 3 nm bandwidth from a down-conversion source is split between two paths using a calcite crystal. Polarizations in individual paths are prepared using optional half-wave plates (HWPs) $H_1$ and $H_2$. Noise $\boldsymbol\Lambda$ is introduced using four sets of orientations of HWPs $H_3$ and $H_4$ and quarter-wave plates $Q_1$ and $Q_2$ mounted on computer-controlled motorized rotation stages. Polarization components of the output state are filtered using a HWP $H_5$ and a second calcite crystal. Interference is realized in the common-path configuration using HWP $H_6$ oriented at $45^\circ$ degrees with respect to the setup plane and a polarizing beam splitter $P$ with output ports monitored by single photon detectors $D_+$ and $D_-$. Two other detectors $D_0$ and $D_1$ are used to determine the overall count rate. The phase shift $\phi$ is introduced by rotating the second calcite crystal about an axis perpendicular to the setup plane using a closed loop piezo actuator. {\bf b}, orientations of principal axes of wave plates $H_3, H_4, Q_1, Q_2$ that realize path-dependent unitaries $\proj[Q]{0} \otimes \hat{K}_{S}^{(0)} + \proj[Q]{1} \otimes \hat{K}_{S}^{(1)}$.}
\label{Fig:Setup}
\end{figure}

{\bf Experiment.}
In order to expose more complex aspects of the trade-off between which-way information and interference visibility for a particle with an internal structure, we investigated experimentally interference of a single photon in a noisy Mach-Zehnder interferometer shown in Fig.~\ref{Fig:Setup}, using photon polarization as the internal subsystem $S$. Single photons with $810$~nm central wavelength were generated by type-II spontaneous parametric down-conversion in a $30$~mm long PPKTP crystal pumped with 14~mW of $405$~nm wavelength light from continuous wave diode laser, and heralded by detection of orthogonally polarized conjugate photons from the same pairs. The photons, after sending through a $3$~nm bandwidth interference filter and transmitting via a single mode fiber, were split between two paths using a calcite displacer. Equal splitting was ensured by using a fiber polarization controller. Removable half-wave plates $H_1$ and $H_2$ were used to prepare any combination of horizontal $h$ and vertical $v$ polarization states for the two paths.

Inside the interferometer the photons were subjected to path-dependent unitaries $\proj[Q]{0} \otimes \hat{K}_{S}^{(0)} + \proj[Q]{1} \otimes \hat{K}_{S}^{(1)}$, where the pair $(\hat{K}_{S}^{(0)} , \hat{K}_{S}^{(1)} )$ was an equally weighted mixture of $(\hat{\openone}, \hat{\openone})$, $(\hat{X}, \hat{X})$, $(\hat{Y}, - \hat{Y})$, and $(\hat{Z}, \hat{Z})$, with $\hat{X}$, $\hat{Y}$, and $\hat{Z}$ denoting Pauli operators. The unitaries were realized by wave plates $H_3, H_4, Q_1, Q_2$ mounted on motorized rotation stages. This procedure scrambled the polarization state in a single arm to the completely mixed state. But the noise had a non-trivial effect on the joint path-polarization state, as
\begin{equation}
{\boldsymbol\Lambda}_{01} (\hat{\sigma}) = \frac{1}{4} \bigl( \hat{\sigma} + \hat{X} \hat{\sigma} \hat{X} - \hat{Y} \hat{\sigma} \hat{Y}
+ \hat{Z} \hat{\sigma} \hat{Z} \bigr)
= \frac{1}{2} \hat\sigma^T,
\end{equation}
which provides a physical realization of Eq.~(\ref{Eq:Lambda01transp}) for $d=2$.
We treat the realized noisy channel as a black box simulating interaction with an environment and analyze the maximum amount of which-way information that might have leaked to the environment in the worst-case scenario. We have seen that another realization of this channel, given in Eq.~(\ref{Eq:ChannelExplicit}), can reveal certain which-way information depending on the preparation of the internal degree of freedom.

In order to estimate which-way information possibly deposited in the environment, we used four different combinations of preparations
for the photon polarization in the upper and the lower paths, denoted jointly as $\mu = hh, hv, vh, vv$.
After the simulated interaction with the environment, we filtered out from each path either horizontal or vertical component using the half-wave plate $H_5$, which directed selected polarizations to the same output port after the second calcite crystal. We used a common-path setup based on the half-wave plate $H_6$ and the polarizer $P$ monitored by single photon detectors $D_\pm$ to measure
the corresponding fractional visibilities $V^{\mu\nu}$.
The index $\nu$ labelling filters also assumes one of four values: $hh$, $hv$, $vh$, or $vv$. The two non-interfering output ports after the second calcite crystal were monitored by detectors $D_0$ and $D_1$ in order to normalize overall count rates. The measured count rates were adjusted by binomial resampling to correct for non-uniform detection efficiencies of the detectors.

\begin{table*}
\begin{tabular}{c}
\\
\\
\\
\\
{\bf a)}
\end{tabular}
\hspace{5mm}
\begin{tabular}{r|cccc}
\diagbox{$\mu$}{$\nu$} & $hh$ & $hv$ & $vh$ & $vv$ \\
\hline
$hh$ & $\frac{1}{2}$ & 0 & 0 & 0 \\
$hv$ & 0 & 0 & $\frac{1}{2}$ & 0 \\
$vh$ & 0 & $\frac{1}{2}$ & 0 & 0 \\
$vv$ & 0 & 0 & 0 & $\frac{1}{2}$
\end{tabular}
\hspace{15mm}
\begin{tabular}{c}
\\
\\
\\
\\
{\bf b)}
\end{tabular}
\hspace{5mm}
\begin{tabular}{c|c|c||c|c|c||}
$\mu\nu$ & $p^{\mu\nu}$ & $V^{\mu\nu}$ & $\mu\nu$ & $p^{\mu\nu}$ & $V^{\mu\nu}$ \\
\hline
$hh,hh$ & $0.489$ & $0.476$ & $hh,vv$ & $0.512$ & $0.104$ \\
$hv,vh$ & $0.513$ & $0.488$ & $hv,hv$ & $0.490$ & $0.039$ \\
$vh,hv$ & $0.511$ & $0.479$ & $vh,vh$ & $0.490$ & $0.032$ \\
$vv,vv$ & $0.489$ & $0.478$ & $vv,hh$ & $0.512$ & $0.100$
\end{tabular}
\caption{{\bf Characterizing noisy interference.} {\bf a}, Theoretical predictions
of fractional visibilities $V^{\mu\nu}$ for preparations $\mu$ and filters $\nu$ chosen in the rectilinear basis of horizontal $h$ and vertical $v$ polarizations. {\bf b}, Experimental values of filtering probabilities $p^{\mu\nu}$ and fractional visibilities $V^{\mu\nu}$. Uncertainties are below $0.003$ for all values shown.}
\label{Tab:FracVis}
\end{table*}

Because individual paths inside the interferometer are completely depolarized by the interaction with the environment, the theoretical probability of filtering out the photon for any selection of the preparation $\mu$ and the filter $\nu$ equals $p^{\mu\nu}=\frac{1}{2}$. Theoretical predictions for fractional visibilities are collected in Tab.~\ref{Tab:FracVis}(a).
It is seen that only few combinations of preparations and filters are expected to produce interference fringes.
Let us select for a moment a single preparation $\mu$. After filtering a pair of polarization components one could use the remaining components to measure also the second fractional visibility for complementary polarizations.
For example, $\nu=vv$ is complementary to $\nu=hh$, and $\nu=vh$ is complementary to $\nu=hv$.
This is an example of a scenario covered by the
inequality (\ref{Eq:VisOrthogonal}). Although the photon is equally likely to choose either of the two complementary filters, interference fringes are visible only for one of them and the right hand side of (\ref{Eq:VisOrthogonal}) is at most $\frac{1}{2}$.

Suppose now that before interaction with the environment the photon polarization is averaged to the completely mixed state and $\hat{\varrho}_0 = \hat{\varrho}_1 = \frac{1}{2}\hat{\openone}$, for example by switching randomly between four preparations. To estimate which-way information available in this case, let us take on the left hand side of Eq.~(\ref{Eq:sumklalphas}) four non-zero coefficients:
\begin{align}
\alpha_{hh,hh}  & = \frac{1}{2} e^{i\theta_1}, &
\alpha_{hv,vh}  & = \frac{1}{2} e^{i\theta_2}, \nonumber \\
\alpha_{vh,hv} & = \frac{1}{2} e^{i\theta_3}, &
\alpha_{vv,vv} & = \frac{1}{2} e^{i\theta_4}
\end{align}
where $\theta_1, \ldots, \theta_4$ are arbitrary phases. The corresponding operator
\begin{multline}
\hat{U} = e^{i\theta_1} \proj{h} \otimes \proj{h} + e^{i\theta_2} \ket{h}\bra{v} \otimes \ket{v} \bra{h} \\
+ e^{i\theta_3} \ket{v} \bra{h} \otimes \ket{h}\bra{v} + e^{i\theta_4} \proj{v} \otimes \proj{v}
\end{multline}
is unitary for any choice of phases $\theta_1, \ldots, \theta_4$. Therefore Eq.~(\ref{Eq:GenVisBound}) provides a bound on generalized visibility in the form $V_G \ge  | e^{i\theta_1} V^{hh,hh} +  e^{i\theta_2} V^{hv,vh} + e^{i\theta_3} V^{vh,hv} + e^{i\theta_4} V^{vv,vv} |/2$. Maximization over phases yields
\begin{multline}
\label{Eq:SwapEstimate}
V_G \ge \frac{1}{2} \bigl( | V^{hh,hh} | + | V^{hv,vh}| + |V^{vh,hv} | + |V^{vv,vv}| \bigr).
\end{multline}
This bound on generalized visibility can be applied directly to experimental data in order to estimate available which-way information for mixed polarization preparation.

\begin{figure}
\includegraphics[width=8.8cm]{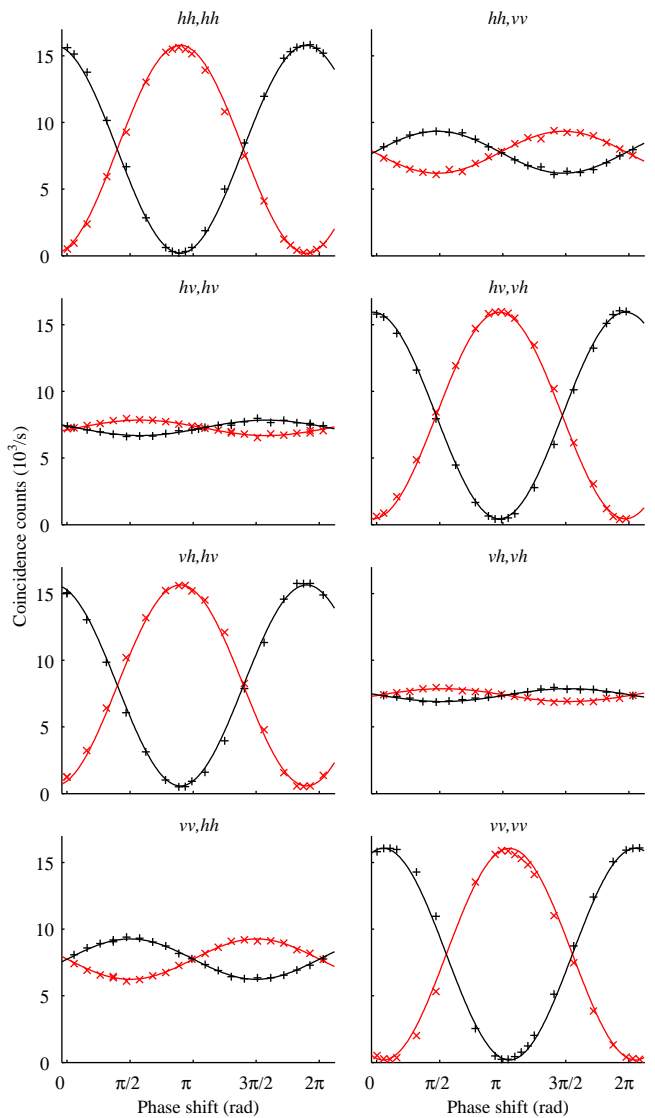}
\caption{{\bf Interference fringes in a noisy interferometer.} Count rates on detectors $D_+$ ($+$, black) and $D_-$ ($\times$, red) conditioned upon detection of conjugate heralding photons from the down-conversion source as a function of the phase shift in the interferometer. Panels are labelled as $\mu,\nu$, where $\mu$ specifies polarization preparation and $\nu$ defines polarizations filtered from the two paths. Solid lines are sinusoidal fits used to determine fractional visibilities. Experimental uncertainties of count rates are smaller than the size of the symbols.}
\label{Fig:Fringes}
\end{figure}

In order to measure fractional visibilities, the phase in the interferometer was adjusted by rotating one of the calcite crystals about an axis perpendicular to the plane of the setup using a closed loop piezo-electric actuator. For each selected phase value we averaged over four unitaries simulating the interaction with the environment.
Experimentally measured interference fringes for combinations of preparations $\mu$ and filters $\nu$ entering the inequality (\ref{Eq:SwapEstimate}) are shown in Fig.~\ref{Fig:Fringes} along with results obtained for complementary filters.
Probabilities $p^{\mu\nu}$ and fractional visibilities $V^{\mu\nu}$ determined from these measurements are collected in Tab.~\ref{Tab:FracVis}(b). Inserting these data in Eq.~(\ref{Eq:SwapEstimate}) gives $V_G \ge 0.960 \pm 0.006$.
This translates into an upper bound on distinguishability in the form $D\le 0.28 \pm 0.02$ according to Eq.~(\ref{Eq:Dlesum}). Remarkably, this stringent bound is determined from just few measurements of fractional visibilities rather than complete quantum process tomography. In contrast, consider an individual pure preparation and complementary projective filters, corresponding to a pair of graphs shown in a single row of
Fig.~\ref{Fig:Fringes}. In this case, experimental results yield at most $V_G \ge 0.580 \pm 0.006$ using Eq.~(\ref{Eq:VisOrthogonal}), which implies
$D\le 0.815 \pm 0.005$. This separation between estimated on distinguishability for mixed and pure polarization preparations strikingly demonstrates the convoluted effects of internal degrees of freedom in a which-way experiment.

\section*{Discussion}

The scenario considered in this work assumed a clear distinction between the environment and the spin that are both external to the path subsystem. Tha amount of available which-way information was determined solely from the quantum state of the environment after the interaction with the particle in the interferometer and no classical information about either spin preparation or measurement results could be used for that purpose. This makes our approach distinct from previous studies of the quantum erasure phenomenon,\cite{BjorkKarlsson,ScullyDruhl,EnglertBergou,KimYuKulik,MaKoflerZeilinger} where postselection carried out on the environment can restore conditional interference fringes. Here the environment was treated as an adversary whose information about the path taken by the particle we try to control and estimate by manipulating the spin subsystem at the preparation and the detection stages.

In this context, the analysis of a which-way experiment with an internal degree of freedom points at non-trivial issues in prepare-and-measure protocols for quantum key distribution over complex noisy channels. Suppose that a cryptographic key is to be established by sending a particle along a randomly selected one of two paths, while the key security is to be verified by preparing occasionally the particle in a superposition of the two paths and measuring its coherence at the output by sampling interference fringes. If the particle is equipped with spin, the trade-off between distinguishability and visibility derived in this work implies that certain eavesdropping strategies require the sender to use randomness in spin preparation in order to ensure security. Further, to verify the security the sender and the receiver may need to combine data collected for an array of spin preparations and filterings.

An entanglement-based analog of the scenario analyzed above would be to prepare jointly two particles in a maximally entangled path-spin state and to subject one of them to interaction with the environment modelling an eavesdropping attempt. In this case, spin can play the role of a ``shield'' subsystem fully protecting the privacy of a cryptographic key generated by detecting particles in individual paths, even though the noise present in the bipartite state may prevent entanglement distillation at the same rate as key generation.\cite{Private1,Private2,Private3,Private4} To complete the parallel,
estimates for generalized visibility determined from fractional visibility measurements can be viewed as a dynamical analog of recently introduced privacy witnesses.\cite{PrivWitness}

\bigskip

\section*{Methods}

Our basic tool will be the Choi-Jamio{\l}kowski isomorphism\cite{Choi,Jamiolkowski} between linear maps and operators acting on the duplicated Hilbert space. Let us define a maximally entangled state of two replicas of the system $QS$ as $\ket[QSQ'S']{\Phi_+} = \ket[QQ']{\Phi_+} \otimes \ket[SS']{\Phi_+}$, where
$\ket[QQ']{\Phi_+} = \bigl( \ket[Q]{0}\ket[Q']{0} + \ket[Q]{1}\ket[Q']{1} \bigr) /\sqrt{2}$ and
$\ket[SS']{\Phi_+} = \bigl( \sum_{l=0}^{d-1} \ket[S]{l} \ket[S']{l}\bigr) /\sqrt{d}$.
The Choi-Jamio{\l}kowski state corresponding to a channel ${\boldsymbol\Lambda}$ is given by
$\hat{\Lambda}_{QSQ'S'} = ({\bf I} \otimes \boldsymbol\Lambda) \bigl( \proj[QSQ'S']{\Phi_+} \bigr)$, where the identity map ${\bf I}$ acts on the subsystem $QS$, while ${\boldsymbol\Lambda}$ acts on $Q'S'$. The transformation of a state $\hat{\varrho}' = {\boldsymbol\Lambda}(\hat{\varrho})$ can be written as
\begin{equation}
\hat{\varrho}'_{Q'S'} = 2d \Tr_{QS} [ \hat{\Lambda}_{QSQ'S'} \hat{\varrho}_{QS}^T]
\end{equation}
where the transposition $^T$ is performed in the same basis in which the state $\ket[QSQ'S']{\Phi_+}$ has been defined.
The input state $\hat{\varrho}$ is encoded in the subsystem $QS$, while the output state is inscribed in the subsystem $Q'S'$.

In order to take into account information deposited in the environment in course of the interaction, we will consider a purification
of the state $\hat{\Lambda}_{QSQ'S'}$:
\begin{equation}
\hat{\Lambda}_{QSQ'S'} = \Tr_{E} \bigl( \proj[QSQ'S'E]{\Lambda} \bigr).
\end{equation}
Because the particle is not transferred between the two paths, the purified state has the form
\begin{equation}
\ket[QSQ'S'E]{\Lambda} = \frac{1}{\sqrt{2}} \bigl( \ket[QQ']{00}\ket[SS'E]{\Lambda_0} + \ket[QQ']{11}\ket[SS'E]{\Lambda_1}\bigr),
\end{equation}
where the states $\ket[SS'E]{\Lambda_i}, i=0,1$, are normalized.

The joint state of the particle and the environment after the interaction can be written as:
\begin{eqnarray}
\hat{\varrho}'_{Q'S'E} & = & 2d \Tr_{QS} \bigl( \proj[QSQ'S'E]{\Lambda}  \hat{\varrho}_{QS}^{T} \bigr)  \nonumber \\
& = & \frac{d}{2} \sum_{i,j=0,1} \ket[Q']{i}\bra{j} \otimes \Tr_{S} [ \ket[SS'E]{\Lambda_i} \bra{\Lambda_j}
(2 \bra{i} \hat{\varrho} \ket{j} )^T_S] \nonumber \\
& &
\end{eqnarray}
In the first expression transposition is performed on subsystems $QS$, while in the second expression only on $S$.
Thus, the environment states correlated with the particle present in one or another path are given after normalization by
$\mbox{$\hat{\varrho}'$}^{(i)}_E = d \Tr_{SS'} [ \ket[SS'E]{\Lambda_i} \bra{\Lambda_i} (\hat{\varrho}_i )^T_S]$, where
$\hat{\varrho}_i = 2 \, \bra[Q]{i}\hat{\varrho}_{QS}\ket[Q]{i}$ are the normalized internal states of the particle in the upper and the lower arms of the interferometer.
Distinguishability can only increase by access to additional subsystems that purify the states. We will take as purifications of
$\mbox{$\hat{\varrho}'$}^{(i)}_{E}$ states $\ket[SS'E]{{\cal E}_i} = \hat{U}_{SS'}^{(i)}\sqrt{d(\hat{\varrho}_i )^T_S}\ket[SS'E]{\Lambda_i}$, where $\hat{U}_{SS'}^{(i)}$ are arbitrary unitaries. For any choice of these unitaries we will have
\begin{equation}
D \le \sqrt{1 - \bigl| \braket[SS'E]{{\cal E}_0}{{\cal E}_1} \bigr|^2}.
\end{equation}
The most stringent bound is obtained by minimizing the right hand side over  $\hat{U}_{SS'}^{(0)}$ and $\hat{U}_{SS'}^{(1)}$. This can be equivalently written as $D \le \sqrt{1-V_G^2}$, where
\begin{equation}
V_G  =   \max_{\hat{U}_{SS'}} \left| \bra[SS'E]{\Lambda_1} \sqrt{d(\hat{\varrho}_1 )^T_S} \hat{U}_{SS'} \sqrt{d(\hat{\varrho}_0 )^T_S}\ket[SS'E]{\Lambda_0}\right|
\end{equation}
and $\hat{U}_{SS'}$ is an arbitrary unitary, or using the trace operation
\begin{equation}
\label{Eq:V=maxUSS'}
V_G = d \max_{\hat{U}_{SS'}} \left|\Tr_{SS'} \left( \hat{U}_{SS'} \sqrt{( \hat{\varrho}_0 )^T_S} \Tr_{E} \bigl( \ket[SS'E]{\Lambda_0} \bra{\Lambda_1} \bigr) \sqrt{(\hat{\varrho}_1 )^T_S} \right)\right|
\end{equation}
The maximum can be expressed in terms of the trace norm as
\begin{equation}
V_G =  d \left|\left| \sqrt{(\hat{\varrho}_0 )^T_S} \Tr_{E} \bigl( \ket[SS'E]{\Lambda_0} \bra{\Lambda_1} \bigr) \sqrt{(\hat{\varrho}_1 )^T_S} \right|\right|.
\end{equation}
The channel characteristics enters the above formula through
\begin{align}
\Tr_{E} \bigl( \ket[SS'E]{\Lambda_0} \bra{\Lambda_1} \bigr) & = 2 \, \bra[QQ']{00} \hat{\Lambda}_{QSQ'S'} \ket[QQ']{11} \nonumber \\
& =  ({\bf I} \otimes {\boldsymbol\Lambda}_{01}) \bigl( \proj[SS']{\Phi_+} \bigr),
\label{Eq:TrE=Lambda01}
\end{align}
where ${\boldsymbol\Lambda}_{01}$ has been defined in Eq.~(\ref{Eq:Lambda01}).
Using the fact that $\bigl( \sqrt{(\hat{\varrho}_i )^T_S} \otimes \hat{\openone}_{S'} \bigr) \ket[SS']{\Phi_+} =
\bigl( \hat{\openone}_{S} \otimes \sqrt{(\hat{\varrho}_i )_{S'}} \bigr) \ket[SS']{\Phi_+}$ yields Eq.~(\ref{Eq:GenVis}).

The expression for fractional visibilities $V^{\mu\nu}$ given in Eq.~(\ref{Eq:FracVis}) can be rewritten in the tensor form
\begin{equation}
\label{Eq:FracVisTensor}
V^{\mu\nu} = d \Tr \bigl\{ \bigl[
\bigl( \ket[S]{\psi_0^{\mu}} \bra{\psi_1^{\mu}} \bigr)^T \otimes \ket[S']{\chi_1^{\nu}} \bra{\chi_0^{\nu}} \bigr]
( {\bf I} \otimes \boldsymbol\Lambda_{01}) \bigl( \proj[SS']{\Phi_+} \bigr)
\bigr\}.
\end{equation}
Maximization in Eq.~(\ref{Eq:V=maxUSS'}) can be extended to all operators $\hat{U}_{SS'}$ satisfying
$\hat{U}_{SS'}^\dagger \hat{U}_{SS'} \le \hat{\openone} \otimes \hat{\openone}$. Therefore for any such operator we have
\begin{equation}
\label{Eq:BoundTensor}
V_G \ge d \left|\Tr_{SS'} \left( \bigl( \sqrt{(\hat{\varrho}_1 )^T_S} \hat{U}_{SS'} \sqrt{( \hat{\varrho}_0 )^T_S} \bigr)
({\bf I} \otimes {\boldsymbol\Lambda}_{01}) \bigl( \proj[SS']{\Phi_+} \bigr) \right)\right|,
\end{equation}
where we used Eq.~(\ref{Eq:TrE=Lambda01}).
Consequently, tracing both sides of Eq.~(\ref{Eq:sumklalphas}) multiplied by $d ( {\bf I} \otimes \boldsymbol\Lambda_{01}) \bigl( \proj{\Phi_+} \bigr)$ and comparing resulting expressions with Eqs.~(\ref{Eq:FracVisTensor}) and (\ref{Eq:BoundTensor}) immediately yields a lower bound on the generalized visibility presented in Eq.~(\ref{Eq:GenVisBound}).

\section*{Acknowledgements}
We wish to acknowledge insightful discussions with B.-G. Englert, P. Raynal, R. Demkowicz-Dobrza\'{n}ski and W. Wasilewski. This research was supported by
the European Union FP7 project Q-ESSENCE (Grant Agreement no.\ 248095), ERA-NET project QUASAR, and the Foundation for Polish Science TEAM project cofinanced by the EU European Regional Development Fund.

\section*{Author contributions}
All authors contributed extensively to the work presented in this paper.

\section*{Competing financial interests}
The authors declare no competing financial interests.

\end{document}